\documentclass[10pt,letterpaper]{article}
\usepackage[top=0.85in,left=2.75in,footskip=0.75in]{geometry}
\usepackage{amsmath,amssymb}
\usepackage{changepage}
\usepackage[utf8x]{inputenc}
\usepackage{textcomp,marvosym}
\usepackage{cite}
\usepackage{nameref,hyperref}
\usepackage[right]{lineno}
\usepackage{microtype}
\DisableLigatures[f]{encoding = *, family = * }
\usepackage[table]{xcolor}
\usepackage{array}
\newcolumntype{+}{!{\vrule width 2pt}}
\newlength\savedwidth

\raggedright
\usepackage[aboveskip=1pt,labelfont=bf,labelsep=period,justification=raggedright,singlelinecheck=off]{caption}

\makeatletter
\renewcommand{\@biblabel}[1]{\quad#1.}
\makeatother

\date{}

\usepackage{lastpage,fancyhdr,graphicx}
\usepackage{epstopdf}
\fancyheadoffset[L]{2.25in}
\fancyfootoffset[L]{2.25in}

\begin{document}
\vspace*{0.2in}

\begin{flushleft}
{\Large
\textbf\newline{The advantage of being slow: the quasi-neutral contact process}
}
\newline
\\
Marcelo Martins de Oliveira\textsuperscript{1\Yinyang},
Ronald Dickman\textsuperscript{2\Yinyang}
\\
\bigskip
\textbf{1} Departamento de F\'{\i}sica e Matem\'atica,
CAP, Universidade Federal de S\~ao Jo\~ao del Rei,
36420-000 Ouro Branco, Minas Gerais - Brazil
\\
\textbf{2} Departamento de F\'{\i}sica and
National Institute of Science and Technology for Complex Systems,
ICEx, Universidade Federal de Minas Gerais,
C. P. 702, 30123-970 Belo Horizonte, Minas Gerais - Brazil

\bigskip

\Yinyang These authors contributed equally to this work.

* mmdeoliveira@ufsj.edu.br

\end{flushleft}

\section*{Abstract}
According to the competitive exclusion principle, in a finite ecosystem, extinction occurs naturally when two or
more species compete for the same resources.
An important question that arises is: when coexistence is not possible, which mechanisms confer an advantage to a given species
against the other(s)? In general, it is expected that the species with the higher reproductive/death ratio will win the competition,
but other mechanisms, such as asymmetry in interspecific competition or unequal diffusion rates, have been found to change this scenario dramatically. In this work, we examine competitive advantage in the context of quasi-neutral population models, including stochastic models
with spatial structure as well as macroscopic (mean-field) descriptions. We employ a two-species contact process in which the ``biological clock" of one species is a factor of $\alpha$ {\em slower} than that of the other species. Our results provide new insights into how stochasticity and competition interact to determine extinction in finite spatial systems. We find that a species with a slower biological clock has an advantage if resources are limited,
winning the competition against a species with a faster clock, in relatively small systems.
Periodic or stochastic environmental variations also favor the slower species,
even in much larger systems.

\section{Introduction}
The search for mechanisms that permit the coexistence and maintenance of species diversity is an important problem in ecology \cite{may}.
In large, well-mixed populations, the dynamics can be accurately described by deterministic rate equations such as the paradigmatic Lotka-Volterra model \cite{may,rockwood,pastor}.
However, deterministic equations cannot describe the number fluctuations and spatial degrees of freedom observed in finite and spatially extended systems \cite{durrett}. Furthermore, there is evidence that spatial structure can facilitate coexistence of similar competitors \cite{condit,coexistence1}. Local symbiotic interspecific interactions can also favor coexistence in these cases \cite{scp1,scp2}.

In a finite ecosystem, extinction arises naturally when two or more species compete for the same resources, as predicted by the {\it competitive
exclusion principle } \cite{levin,meerson10, compexclusion}.
An important question that arises is: When coexistence is not possible, which mechanisms confer advantage to a given species against the others?
In general it is expected that the species with the higher reproductive/death ratio will win the competition \cite{coexistence2}, but other
mechanisms have been found to change this scenario dramatically. For example,  Gabel, Meerson and Redner \cite{ss} showed recently that asymmetry in interspecific competition can induce survival of the scarcer species (SS), allowing it to out-compete a more populous one. The same conclusion was shown to apply to spatially extended model \cite{sss}. Pigolotti and Benzi \cite{diffast}
demonstrated that an effective selective advantage emerges when two species diffuse at different rates; in this case, competition is biased towards the fastest
species. In addition to two-species competition, such behaviors can also arise in social systems, for example, in opinion formation \cite{crokidakis}.

In this work we examine competitive advantage in the context of stochastic spatial population models, using a two-species contact process. Originally proposed as a toy model for epidemic spreading, the contact process (CP) \cite{harris-CP} can also be interpreted as a stochastic single-species birth-and-death process with spatial structure \cite{durrett}.
In the CP, each individual can reproduce asexually with rate $\lambda$, or die with a unitary rate.
Reproduction is only possible at vacant sites neighboring the parent organism, thus representing competition for space.
As the reproduction rate $\lambda$ is varied, the system undergoes a continuous phase
transition between extinction and survival at a critical value, $\lambda_c$ \cite{marro,odor07,henkel,hinrichsen,odor04}.
In the two-species CP considered here, both species have the same reproduction/death ratio, so that in isolation each would have
the same stationary population density.  The two species differ in their overall rapidity: the reproduction and death rates
of the slow species ($B$) are both a factor of $\alpha < 1$ smaller than those of the fast species ($A$).
Thus for each generation of the slower species, $1/\alpha$ generations will have transpired for the faster species.
In evolutionary ecology such a situation is called ``quasi-neutral", since both species are only neutral\footnote{Neutral theories, in the context of ecology and genetics, assume that random processes, such as demographic stochasticity, dispersal,
speciation and ecological drift, have a stronger impact than differences in species and their
interdependence \cite{neutral2,neutral3,neutral4,neutral1}.} in the deterministic limit.
More recently, Kogan et al. \cite{kogan} applied a perturbation method based on time-scale separation to study quasi-neutral competition in two-strain diseases. They found that the slow strain has a competitive advantage for uniform initial conditions, but the fast strain is more likely to win when a few infectives of both strains are introduced into a susceptible population.

Multispecies (or multitype) contact processes have been used to model systems of sessile species with neutral community structure. They have proven useful in understanding abundance distributions and species-area relationships \cite{weitz, munoz}, but, to our knowledge,  the effects
of quasi-neutral competition have not been addressed yet.

In the present work, we find that the species with the longest lifecycle (i.e., slower dynamics) has a competitive advantage
in an ecosystem with spatial structure when
the resources are finite. We find that environmental fluctuations also confer an advantage
to the slower species.

In a broader perspective, our work highlights the role of demographic and environmental noise when two stochastic populations competing
in the same domain evolve on different timescales. Such a timescale separation can also be relevant in emerging phenomena
in opinion dynamics \cite{boguna}.

The remainder of this paper is organized as follows. In the next
section we introduce the model and analyze its mean-field theory.
In Sec. 3 we present and discuss our results; Sec. 4 is
devoted to conclusions.

\section{Materials and methods}
\subsection*{Model and mean-field theory}

We define the contact process with a faster species (CPFS) as follows. Consider two species, $A$ and $B$, which inhabit in the same lattice. Each individual of species $A$  attempts to create a new individual at one of its first neighbor sites with rate $\lambda$ and dies with rate 1.  Individuals of species $B$  follow the same dynamics but with creation rate $\alpha\lambda$ and death rate $\alpha$, with $0\leq\alpha\leq1$. Note that the dynamics of isolated $A$ or $B$ populations is the same as in the basic contact process, but that the ``biological clock" of species $B$ is $\alpha$ times {\em slower} than that of species $A$. In other words, for each $B$ generation, $1/\alpha$ generations will transpire for species $A$. Each site of the lattice can be in one of three states: empty, occupied by A, or occupied by B. Since simultaneous occupation by A and B is forbidden, the species compete for space.

The macroscopic mean-field (MF) equations for the model are
\begin{eqnarray}
\frac{d\rho_A}{dt}=\lambda(1-\rho_A-\rho_B)\rho_A-\rho_A, \nonumber \\
\frac{d\rho_B}{dt}=\alpha\lambda(1-\rho_A-\rho_B)\rho_B-\alpha\rho_B
\label{mfteqs}
\end{eqnarray}

\noindent which have the stationary solutions $(\rho_A,\rho_B)=(0,0),(0,1-\lambda^{-1})$ and $(1-\lambda^{-1},0)$. In fact, there is a line of stationary solutions, $\rho_A = x \rho_s$, $\rho_B=(1-x)\rho_s,$ where $\rho_s = 1-1/\lambda$ and $0 \leq x \leq 1$.  In a linear stability analysis, there is a zero eigenvalue along this line.  Thus in the presence of noise we expect fluctuations to take the system to one of the absorbing points, $x=1$ or $x=0$.

Neuhauser \cite{neuhauser} proved rigorously that coexistence of competing contact processes (with $\alpha=1$) is impossible on the square lattice, $Z^2$ if the species have equal death rates, and have the same dispersal distribution\footnote{This applies applies only to the case $\lambda_A = \lambda_B$
{\it and} $\mu_A = \mu_B$, and equal dispersal distributions. If one species disperses faster than other than coexistence is possible for some values of $\lambda_A$ and $\lambda_B$.}
This is in agreement with the competitive exclusion principle \cite{levin}, which states that the number of coexisting species is smaller than the number of resources they compete for. (For instance, in our two-species model we have one resource, empty sites, with density $\rho_0=1-\rho_A-\rho_B$.) If both species are supercritical, i.e., have $\lambda_i>\lambda_c$, where $\lambda_c$ is the critical value of the contact process, the species with higher birth rate will win the competition.

In general, it is expected that the superior competitor (the species with the higher reproductive/death ratio) will win the competition \cite{neuhauser}. The model we consider here represents a less obvious scenario, since both species have the same ratio of reproduction to death rates, and in the macroscopic limit, the two species might be expected have the same extinction probability.  However, the species with the fast ``biological clock" is more subject to temporal density fluctuations; the discrete and spatial character of the individuals may change the scenario. To investigate these effects, in the next section we study the CPFS model in finite systems with spatial structure. In finite systems, extinction is always inevitable, due to the existence of absorbing states. In this context, the quasistationary (QS) distribution describes the asymptotic (long-time) properties of a finite system conditioned on survival \cite{yaglom47,ferrari,nasell01,bart}.
The quasi-stationary properties converge to the stationary properties in the limit of infinite system size.

\section{Results and Discussion}

\subsection{Initial-condition dependence}

Before studying finite, stochastic populations, we take a closer look at the MF equations, (\ref{mfteqs}).  Although, as noted, these equations admit stationary solutions dominated by either species, near the critical point $\lambda_c=1$, most of the space of initial conditions (ICs) flows
to steady states with $\overline{\rho}_B > \overline{\rho}_A$.
 Note that Eqs.~(\ref{mfteqs}) imply that

\begin{equation}
\frac{d \ln \rho_B}{dt} = \alpha \frac{d \ln \rho_A}{dt},
\label{dlns}
\end{equation}

\noindent from which one has

\begin{equation}
\rho_B(t) = \rho_{B,0} \left( \frac{\rho_A (t)}{\rho_{A,0}} \right)^\alpha \, .
\label{rhoBt}
\end{equation}

\noindent Then, using the fact that in the active phase, the stationary populations satisfy $\overline{\rho}_A + \overline{\rho}_B = 1- \lambda^{-1}$,
we see that the initial B population density, $\rho_{B,0}^e$, required for equal {\it final} densities is,

\begin{equation}
\rho_{B,0}^e = \rho_{A,0}^\alpha \left( \frac{1 \!-\! \lambda}{2} \right)^{1-\alpha}.
\label{rhoB0e}
\end{equation}

\begin{figure}[!hbt]
\includegraphics[clip,angle=0,width=.8\hsize]{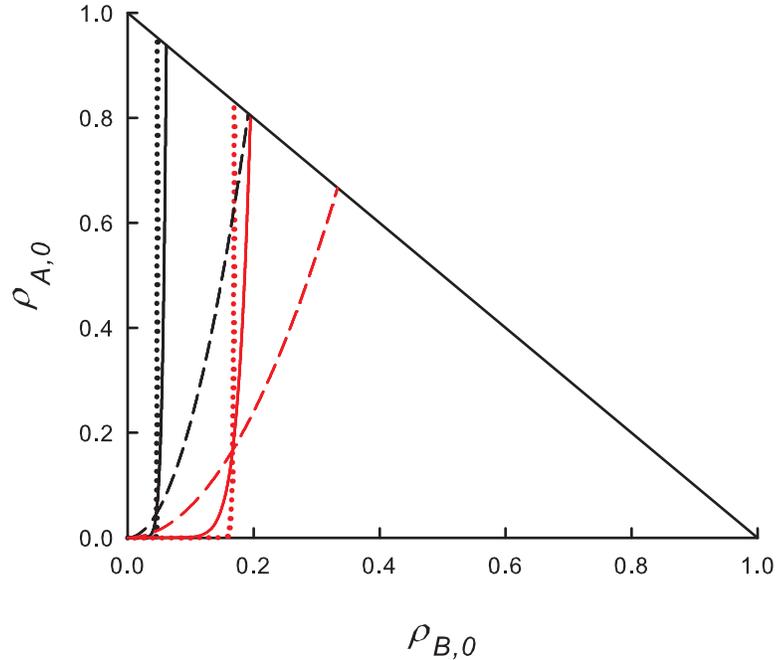}
\caption{\footnotesize{(Color online) MF equations: boundaries between initial populations that lead to
a stationary solution with $\rho_A > \rho_B$ (to the left of the boundary) or vice-versa.
Black curves: $\lambda = 1.1$, red curves: $\lambda = 1.5$.  Dotted lines: $\alpha = 0.01$; solid lines:
$\alpha = 0.1$; dashed lines: $\alpha = 0.5$.
}}
\label{bdrnn}
\end{figure}

\noindent Thus, for $\alpha \ll 1$, $\rho_{B,0}^e$ grows quite slowly with $\rho_{A,0}$, and most ICs yield final states
with $\overline{\rho}_B > \overline{\rho}_A$.
Figure \ref{bdrnn}, shows the separatrix $\rho_{B,0}^e (\rho_{A,0}$ between A- and B-dominated final states for several
choices of $\lambda$ and $\alpha$. The share of ICs leading to $\rho_B > \rho_A$ decreases with increasing $\alpha$ and $\lambda$.

\begin{figure}[!hbt]
\includegraphics[clip,angle=0,width=.9\hsize]{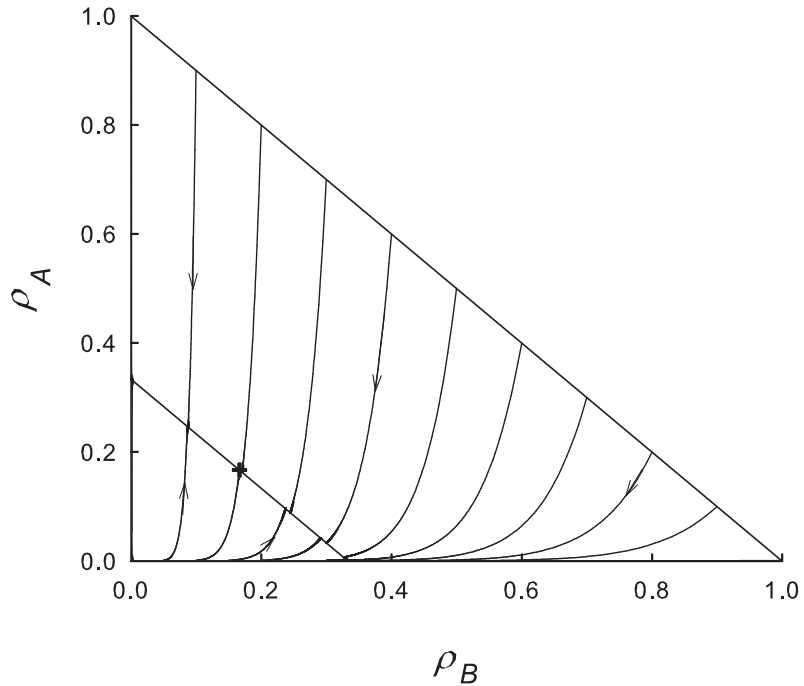}
\caption{\footnotesize{MF equations: trajectories in the $\rho_B$ - $\rho_A$ plane for $\lambda=1.5$ and
$\alpha=0.1$.  All trajectories terminate on the stationary line, $\rho_B + \rho_A = 1/3$.  Most of the
trajectories originating below this line terminate with $\rho_A > \rho_B$, i.e., to the left
of the point (1/6,1/6) denoted by a $\Large{+}$ sign.
}}
\label{traj151}
\end{figure}

Some insight into this apparent advantage of the slow species is afforded by noting that for $\lambda $ close to, but larger than $\lambda_c$,
most ICs lie above the stationary line $\rho_A + \rho_B = 1 - \lambda^{-1}$.  Since the rate of decrease of ln$\rho_B$ is a
factor of $\alpha$ smaller than that of ln$\rho_A$, most ICs in this region flow to a stationary solution with $\rho_B > \rho_A$.
By contrast, the majority of ICs {\it below} the stationary line flow to final states with $\rho_B < \rho_A$.  The more rapid
growth of $\ln \rho_A$ in this region leads to dominance of the fast species, even for some ICs with $\rho_{B,0} >> \rho_{A,0}$.
These trends are illustrated in Fig.~\ref{traj151}.
Summarizing, near the critical point, high initial densities favor the slow species and
{\it vice-versa}.  ICs above the stationary line arise if $\lambda$ is reduced from a value well above $\lambda_c$ to the vicinity
of the critical point, as may occur, for instance, under periodic modulation of $\lambda$ (see Sec. 3.3).

\subsection{MF equations with noise}

\begin{figure}[!hbt]
\includegraphics[clip,angle=0,width=1.\hsize]{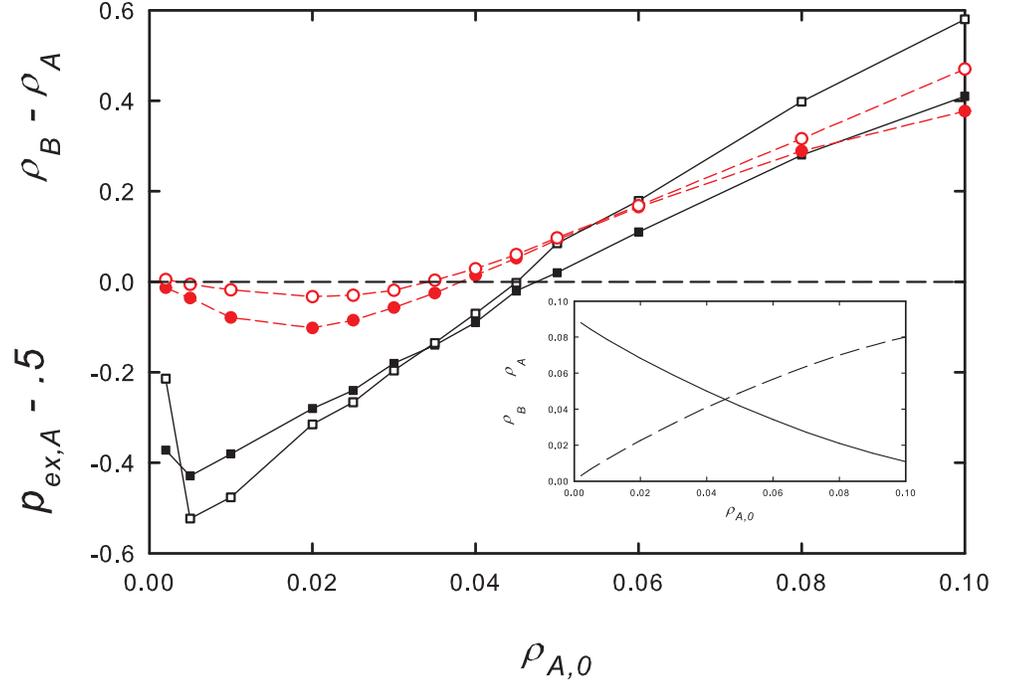}
\caption{\footnotesize{(Color online) MF equations with additive noise.  Species-$A$ extinction probability
$p_{ex,A} - .5$ (filled symbols) and excess mean population density $\rho_B - \rho_A$ (open symbols) versus initial density $\rho_A = \rho_B$ for
$\lambda = 1.1$ and $\alpha = 0.1$.  Noise intensity $\gamma = 0.01$ (squares and solid lines) and 0.1 (circles and dashed lines).
Points reflect averages over $10^3$ - $10^4$ independent realizations.  Density differences are multiplied by a factor of 10
for visibility.
Error bars smaller than symbols.
Inset: stationary population densities $\overline{\rho}_A$ (solid line) and $\overline{\rho}_B$ (dashed line) in the
absence of noise.
}}
\label{pexc}
\end{figure}

Adding noise to the MF equations does not appear to change the above conclusions.  We study the effect of noise by
including independent additive noise terms, $\xi_A(t)$ and $\xi_B (t)$ to Eq.~\ref{mfteqs}, so that
$d\rho_A/dt = \lambda(1-\rho_A-\rho_B)\rho_A-\rho_A + \xi_A(t)$, and similarly for species $B$, where $\xi_A(t)$ is uniform on the interval $[-\gamma, \gamma]$,
and chosen independently at each time step ($dt = 0.01$) of the numerical integration.  Since noise leads to one or another
of the species going extinct, we analyze the extinction probability of species $A$, $p_{ex,A}$, as well as the mean densities
over the evolution, as a function of the initial density,  $\rho_{A,0} = \rho_{B,0}$.
We find that, as in the noise-free case, higher initial densities lead to larger mean densities of the slow species.
Associated with $\rho_B > \rho_A$ is a higher probability of the fast species going extinct first, i.e., $p_{ex,A} > 1/2$.
We verify this for noise intensity $\gamma = 0.01$ - 1, $\lambda = 1.1 - 1,5$, and for $\alpha = 0.1$ - 0.5.
For strong noise (i.e., $\gamma \simeq 1$), however,
the mean time to extinction becomes short (on the order of 10 time units or less), so that the results no longer represent a quasistationary regime.
Figure~\ref{pexc} shows that the species with smaller
mean density has a higher extinction probability.  We have not found evidence for SS (corresponding to $p_{ex,A} > 1/2$ while
$\rho_B < \rho_A$) in the noise-driven MF equations.

\begin{figure}[!hbt]
\includegraphics[clip,angle=0,width=1.0\hsize]{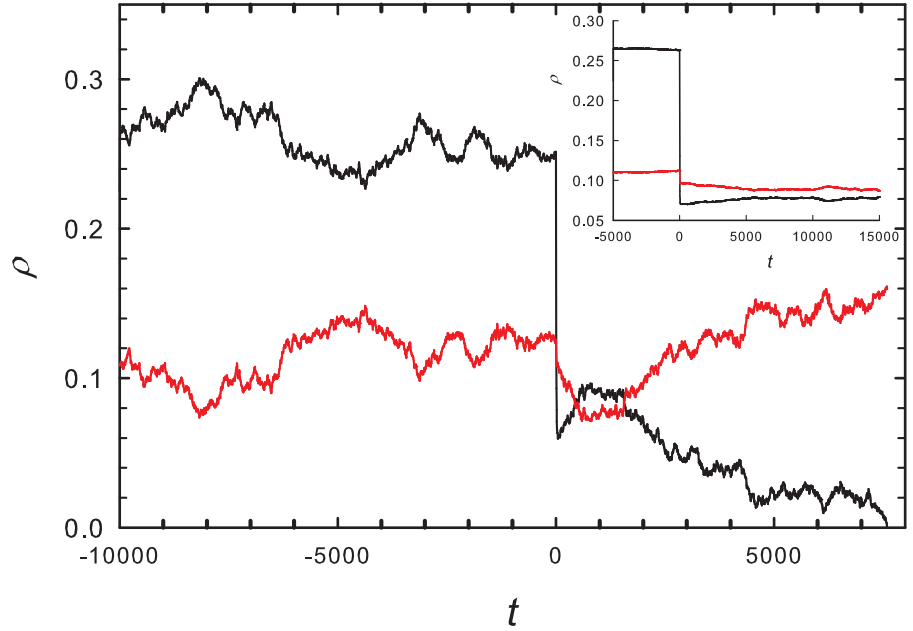}
\caption{\footnotesize{MF equations with additive noise.  Main graph shows the densities of the fast
and slow species (black and red curves, respectively) versus time; at time zero the reproduction rate $\lambda$ is reduced from
1.6 to 1.2.  Other parameters: $\alpha = 0.1$, $\gamma = 0.01$. 
Inset: average densities in a sample of 500 independent realizations.
}}
\label{tsrl1}
\end{figure}

 As noted in the preceding subsection, the slow species should enjoy an advantage when, starting from a stationary state in the
active phase, the creation rate $\lambda$ is reduced.  We verify this prediction in simulations of the noisy MF equations. First, a stationary
state is established for creation rate $\lambda$; then at time zero, this rate is reduced to $\lambda'$.  Figure \ref{tsrl1}, for $\lambda=1.6$,
$\lambda'=1.2$, $\alpha=0.1$ and noise intensity $\gamma = 0.01$, shows that following the sudden decrease in creation rate, the slow
species population is greater on average than that of the fast species, although the latter dominates prior to the reduction.  Similar
results are found for other parameter sets.  As might be expected, following a sudden {\it increase} in the creation rate, the faster species is found to attain a higher density than the slower one.

\subsection{Periodic environment}

A periodic environment is pertinent, for example, to species subject to
seasonal variation of resources or to a parasite subject to the diurnal variation of the host.
We examine the mean-field equations, (\ref{mfteqs}), under periodic forcing of the form
$\lambda = \lambda(t) = \lambda_0 + r \cos \omega t$.
Integrating the equations numerically, we find that, following a transient, the solutions assume a periodic form with the same angular
frequency, $\omega$, as $\lambda(t)$, provided the mean value $\lambda_0 > 1$, the mean-field critical value for constant $\lambda$.
(For $\lambda_0 < 1$ the densities oscillate with an exponentially decaying envelope, while for $\lambda_0 = 1$ the envelope decays
$\propto 1/t$.)

\begin{figure}[!hbt]
\includegraphics[clip,angle=0,width=.98\hsize]{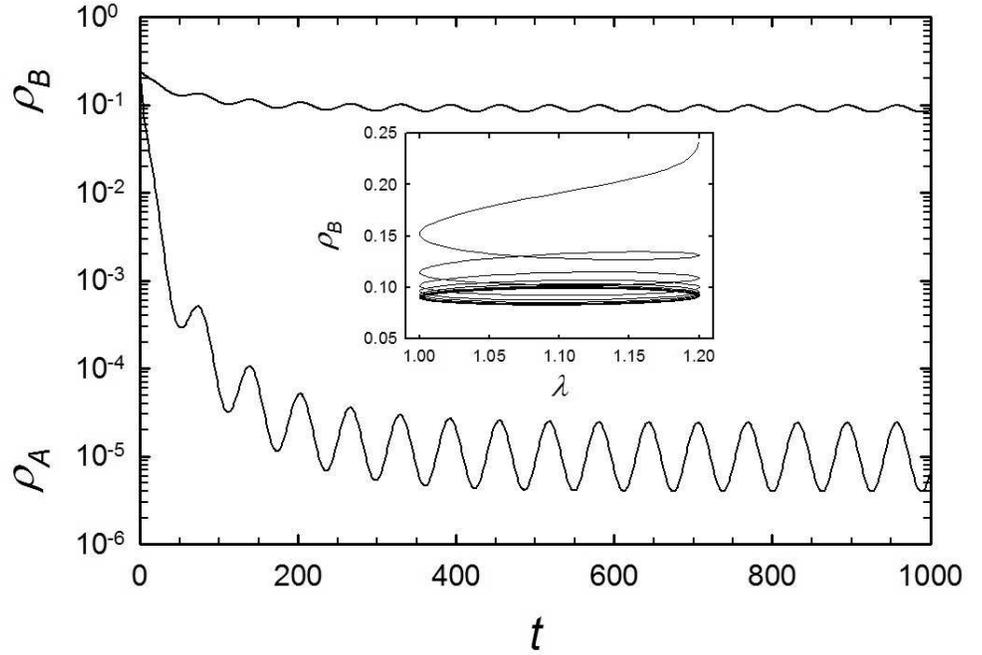}
\caption{\footnotesize{MF equations with periodic $\lambda$: population densities $\rho_B$ (upper) and $\rho_A$ (lower), versus time,
for $\lambda_0 = 1.1$, $r=0.1$, $\omega = 0.1$ and $\alpha=0.1$.  The inset shows $\rho_B (t)$ versus $\lambda(t)$, with an initial
transient followed by a periodic orbit.
}}
\label{rhota}
\end{figure}

\begin{figure}[!hbt]
\includegraphics[clip,angle=0,width=.98\hsize]{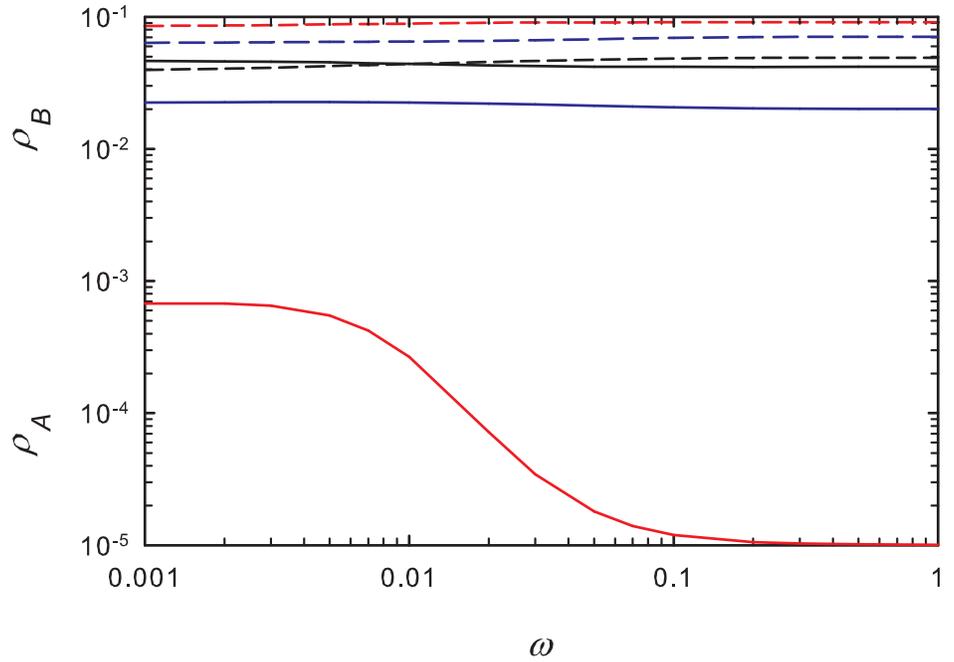}
\caption{\footnotesize{MF equations with periodic $\lambda$: mean densities $\rho_B$ (dashed lines) and
$\rho_A$ (solid) versus $\omega$, for $\lambda_0 = 1.1$ and $r=0.1$.  Red: $\alpha = 0.1$ and $\rho_{A,0} = \rho_{B,0} = 0.25$;
black: $\alpha = 0.1$ and $\rho_{A,0} = \rho_{B,0} = 0.05$;
blue: $\alpha = 0.5$ and $\rho_{A,0} = \rho_{B,0} = 0.25$.  Population densities for $\alpha = 0.5$ and $\rho_{A,0} = \rho_{B,0} = 0.05$
(not shown) are very similar to those for $\alpha = 0.1$ and $\rho_{A,0} = \rho_{B,0} = 0.05$.
}}
\label{mpa}
\end{figure}

\begin{figure}[!hbt]
\includegraphics[clip,angle=0,width=.9\hsize]{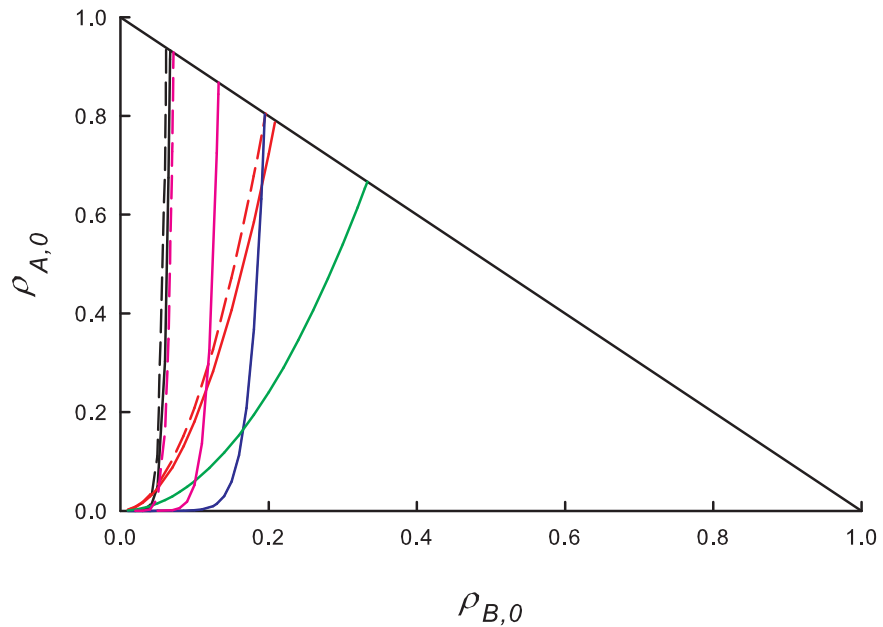}
\caption{\footnotesize{(Color online) Mean-field analysis for periodic environment.  Initial conditions
leading to $\overline{\rho}_A > \overline{\rho}_B$ ($\overline{\rho}_B > \overline{\rho}_A$) correspond to regions to the
left (right) of curves.  Black: $\lambda=1.1$, $\alpha = 0.1$; red: $\lambda=1.1$, $\alpha = 0.5$;
blue: $\lambda=1.5$, $\alpha = 0.1$; green: $\lambda=1.5$, $\alpha = 0.5$; pink: $\lambda=1.1$, $\alpha = 0.1$, but
$r=0.5$. Amplitude of periodic modulation $r=0.1$ in all other cases.
Solid lines: frequency $\omega = 0.01$; dashed lines $\omega = 0.1$. For $\lambda=1.5$ (blue and green) the curves
for the two frequencies are indistinguishable.  
}}
\label{bdra}
\end{figure}

Figure \ref{rhota} shows a typical time series of the population densities,
for $\lambda_0 = 1.1$, $r=0.1$, and $\alpha = \omega = 0.1$; in this case the stationary mean population of the fast species, $\overline{\rho}_A$, is about four orders of magnitude smaller than that of the slow species.  The inset shows that the density of species $B$ is approximately 90$^{\,{\rm o}}$ out of phase with $\lambda$.
The results in Fig.~\ref{rhota} were obtained using equal initial densities, $\rho_{A,0} = \rho_{B,0} = 0.25$.
For these initial densities, we find that $\overline{\rho}_B \gg \overline{\rho}_A$ for a wide range of frequencies, as shown in Fig.~\ref{mpa}.
Varying the initial densities, the system of equations relaxes to a long-time periodic regime with
$\overline{\rho}_B + \overline{\rho}_A = const.$, similar to the stationary line, $\rho_A + \rho_B = 1 - \lambda^{-1}$,
found in MFT for a steady environment.  The slow species enjoys an advantage for large initial densities, smaller $\alpha$ and higher
frequencies.
Fig.~\ref{bdra} shows the lines in the plane of initial densities separating
steady states with $\overline{\rho}_B > \overline{\rho}_A$ and vice-versa.  Similar to the MFT results for constant $\lambda$
discussed above, for small $\alpha$ and $\lambda$ near unity, species $A$ dominates only for ICs in a small region
in this plane, characterized by small initial densities, or by $\rho_{B,0} \ll \rho_{A,0}$.  Increasing either $\alpha$ or $\lambda$, the
advantage of the slow species is reduced.  The advantage of the slow species appears to be robust to increases in the
amplitude $r$ of periodic modulation.

\begin{figure}[!hbt]
\includegraphics[clip,angle=0,width=.95\hsize]{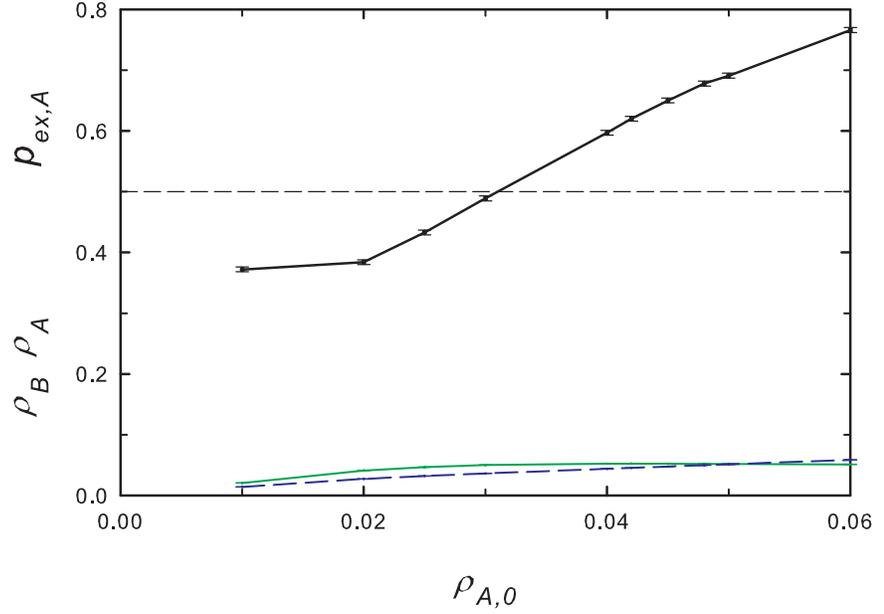}
\caption{\footnotesize{(Color online) Mean-field analysis for periodic environment with independent additive
noise terms affecting each species.  Slow-species extinction probability $p_{ex,A}$ (upper curve)
and mean population densities (lower curves)
$\overline{\rho}_A$ (solid) and $\overline{\rho}_B$ (dashed), versus initial density, with $\rho_{B,0} = \rho_{A,0}$.
Noise intensity $\gamma = 0.1$,
other parameters as in Fig.~\ref{rhota}.
}}
\label{pexa}
\end{figure}

The above conclusions are robust to the inclusion of an additive noise in $\lambda$, corresponding to a
{\it multiplicative} noise in the evolution of $\rho_A$ and $\rho_B$ that affects the two species equally.  Specifically,
we consider $\lambda = \lambda(t) = \lambda_0 + r \cos \omega t + \xi(t)$, with $\xi(t) $ uniform on the interval $[-\gamma, \gamma]$.
(The value of $\xi$ is chosen anew at each time step, $\Delta t = 0.01$.)  In this case the species-dominance patterns identified above
are preserved, even for rather large noise amplitudes ($\gamma = 1$).  If we instead include independent, identically distributed
additive noises $\xi_A(t)$ and $\xi_B(t)$ in the MF equations (\ref{mfteqs}), one or another of the populations eventually goes extinct,
as found for constant $\lambda$.   The behavior of $p_{ex,A}$ is
broadly similar to that found in fixed-$\lambda$ case: for small initial densities,
which lead to $\overline{\rho}_A > \overline{\rho}_B$, we find $p_{ex,A} < 1/2$ (i.e., the slow species
tends to go extinct first).  (Here, the initial densities are equal, and the mean densities
$\overline{\rho}_i$ are time averages from time zero until one of the species goes extinct.)
Figure \ref{pexa} shows, nevertheless, that there is a considerable range of initial densities exhibiting SS, that is, $\overline{\rho}_A > \overline{\rho}_B$ whilst $p_{ex,A} > 1/2$.

Summarizing, in the context of large, well-mixed populations subject to a periodic (sinusoidal) environmental variation,
the slower species generally enjoys an advantage, although situations dominated by the fast species also exist.
A similar pattern is observed under additive noise (independently affecting the two species), including a regime in which the slower species is less populous, but tends to outlive the faster one.

\subsection{Finite stochastic systems}

\subsubsection{Complete Graph}

A {\em complete graph} is one in which all sites are neighbors. Since all sites are equivalent, when formulated on such
a structure, a stochastic process is specified by one or a
few variables, thereby permitting an exact analysis. Since each site interacts with all others, the behavior is mean-field-like.

The state of the CPFS on a complete graph is specified by the variables $n_A$ and $n_B$, the number of sites occupied by species $A$ and $B$, respectively.
On a graph of $N$ sites, the nonzero transition rates $W(n_A',n_B';n_A,n_B)$ (with the primed variables denoting the
state after the transition) are,
\begin{eqnarray}
W(n_A-1,n_B;n_A,n_B)=n_A \nonumber \\
W(n_A,n_B-1;n_A,n_B)=\alpha n_B \nonumber \\
W(n_A+1,n_B;n_A,n_B)=\lambda\frac{n_A}{N}(N -n_A-n_B) \nonumber \\
W(n_A,n_B+1;n_A,n_B)=\alpha\lambda\frac{n_B}{N}(N -n_A-n_B) \nonumber \\
\end{eqnarray}
with $n_A=n_B=0$ absorbing (Note that the subspaces $n_A=0$ and $n_B=0$ are also absorbing).

In order to evaluate the QS probability distribution, we apply an iterative scheme, proposed in \cite{iterative}.
We consider the QS state as the subspace with at least one individual of each species. From any nonabsorbing initial configuration the conditional probability distribution converges to the QS probability distribution  $Q(n_A,n_B)$ after a few iterations. Using the QS distribution, we calculate the QS densities of species $A$ and $B$.
Another important quantity is the lifetime of the QS state, $\tau$, given by $\tau=1/A_0$, with $A_0=\sum_{n_B} Q(1,n_B) W(0,n_B;1,n_B) + \sum_{n_A} Q(n_A,1)W(n_A,0;n_A,1)$ the flux of probability to the absorbing subspace.   Note that $A_{0,A} \equiv \sum_{n_B} Q(1,n_B) W(0,n_B;1,n_B)$ is the exit rate from the QS state to the subspace with $n_A= 0$, and similarly for $A_{0,B} \equiv \sum_{n_A} Q(n_A,1)W(n_A,0;n_A,1)$.  We therefore interpret $1/A_{0,A}$ and $1/A_{0,B}$ as the QS lifetimes of species $A$ and $B$, respectively.

\begin{figure}[!hbt]
\includegraphics[clip,angle=0,width=.9\hsize]{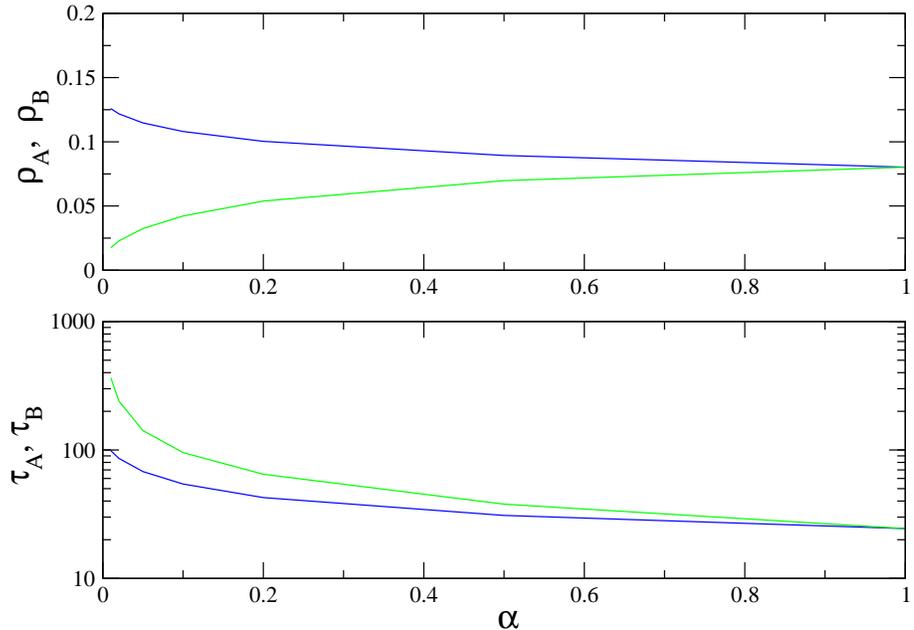}
\caption{\footnotesize{(Color online) Quasistationary population densities of species A (blue) and B (green) and lifetime of the QS state
on a complete graph versus $\alpha$, for $\lambda=1.2$ and $N=200$.
}}
\label{alphaGC}
\end{figure}

In Fig.~\ref{alphaGC} we show the QS densities of species $A$ and $B$ and the lifetime of the QS state as a function of $\alpha$ in the supercritical regime, $\lambda=1.2$, for a graph of $N=200$ sites. Species $A$ (the faster) is always more populous, however, its lifetime is shorter. Therefore, in this case, species $B$, with a smaller population than $A$, is more likely to survive. The advantage of species $B$ vanishes when $\alpha\to 1$, as expected.

\begin{figure}[!hbt]
\includegraphics[clip,angle=0,width=.9\hsize]{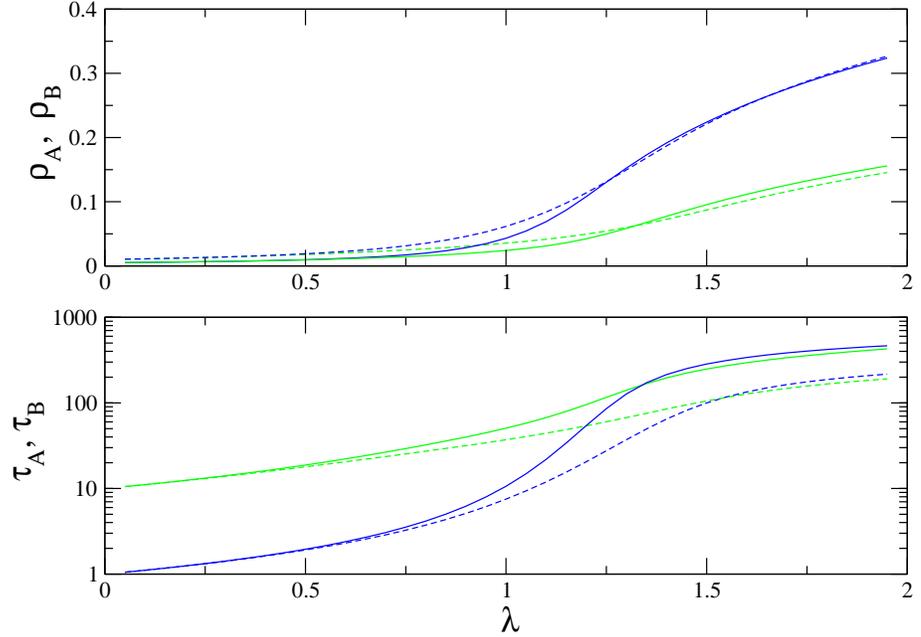}
\caption{\footnotesize{(Color online) Quasistationary population densities (upper panel) and
lifetime of the QS state (lower panel) on a complete graph for species $A$ (blue) and $B$ (green) versus $\lambda$, for $\alpha=0.1$. Graph sizes: $N=100$ (dashed) and $N=200$
}}
\label{lambdaGC}
\end{figure}

In Fig.~\ref{lambdaGC} we show the QS densities of species $A$ and $B$ and the lifetime of the QS state as function of $\lambda$ for $\alpha=0.1$, for graphs of $N=100$ and $N=200$ sites. The slower species $B$ is again less numerous than species $A$ for all values of $\lambda$. There is
nevertheless a range of $\lambda$ values above the critical value $\lambda_c=1$ in which species $B$ survives for a longer time than $A$; increasing $\lambda$ further, the advantage vanishes.

\begin{figure}[!hbt]
\includegraphics[clip,angle=0,width=1.0\hsize]{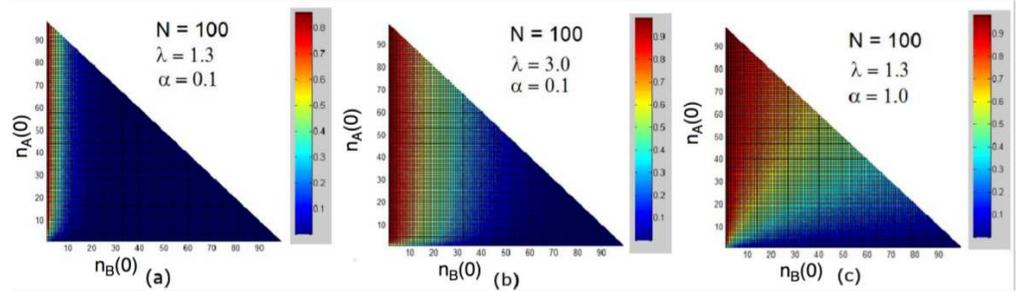}
\caption{\footnotesize{(Color online) Probability of the faster species ($A$) winning the competition for the process on a complete graph, as function of initial species populations, $n_A(0)$ and $n_B(0)$.  Graph size: $N=100$.
}}
\label{survGC}
\end{figure}

The advantage of the slower species in an environment with limited resources is evident in Fig.~\ref{survGC}, which shows the probability of the faster species ($A$) winning the competition. In Fig.~\ref{survGC}(a), for $\lambda$ close to $\lambda_c$, the faster species goes extinct first
for almost all values of the initial densities, similar to the results for the MF equations with additive noise.
The advantage is less pronounced when increasing $\lambda$, as shown Fig.~\ref{survGC}(b), and vanishes when $\alpha=1$ (see Fig.~\ref{survGC}(c)).

Our results are in agreement with the theoretical predictions of Lin {\it et al.} \cite{lin2012}, who showed that demographic fluctuations break the degeneracy displayed by the deterministic rate equation
description of the dynamics of two competing species differing only in the time scales of
their life cycles, and with the results of Kogan {\it et. al.} \cite{kogan} for multi-strain diseases. Therefore the slower species enjoys
a slight competitive advantage over the faster one.

\subsubsection{Exact QS state for small rings}

\begin{figure}[!hbt]
\includegraphics[clip,angle=0,width=.95\hsize]{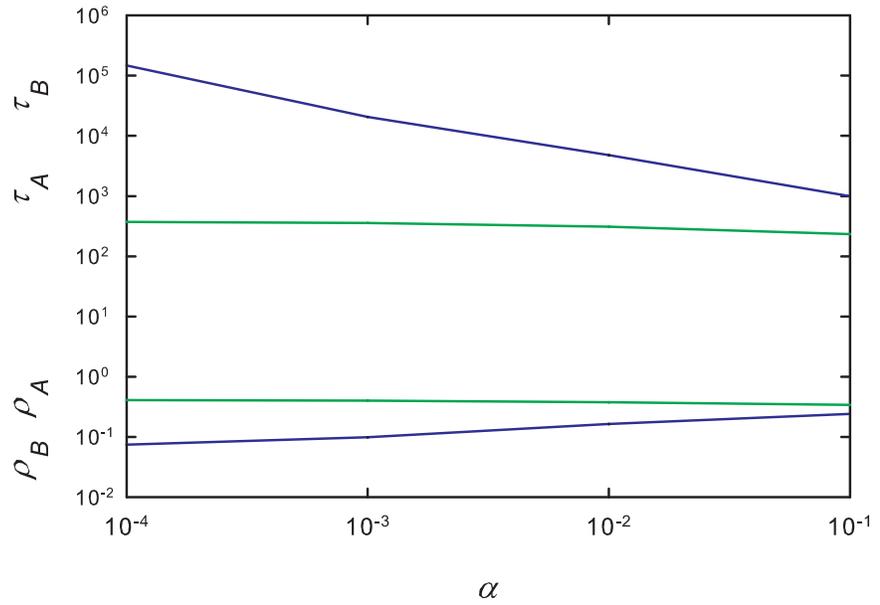}
\caption{\footnotesize{(Color online) Quasistationary population densities (lower) and lifetimes (upper) for fast (green lines) and slow (blue lines)
species versus $\alpha$ on a ring of 14 sites, for $\lambda = \lambda_c$ = 3.29785.
}}
\label{cpfsring}
\end{figure}

Using the iterative method developed in \cite{iterative}, we determined the exact (numerical) QS probability distribution on
rings of up to $L=14$ sites.  Similar to the results for the complete graph, we find that while the QS population of the slow species
is smaller than than that of the fast one, its lifetime is considerably greater.  An example of these results is shown in Fig.~\ref{cpfsring}.
Due to the limited range of system sizes accessible to this analysis, we cannot extrapolate reliably to the infinite-size limit. We
nevertheless note that the differences between densities $\rho_A$ and $\rho_B$, and between lifetimes $\tau_A$ and $\tau_B$, decrease
with increasing system size, consistent with vanishing differences for $L \to \infty$.

\subsubsection{Two dimensions}

\begin{figure}[!hbt]
\includegraphics[clip,angle=0,width=.95\hsize]{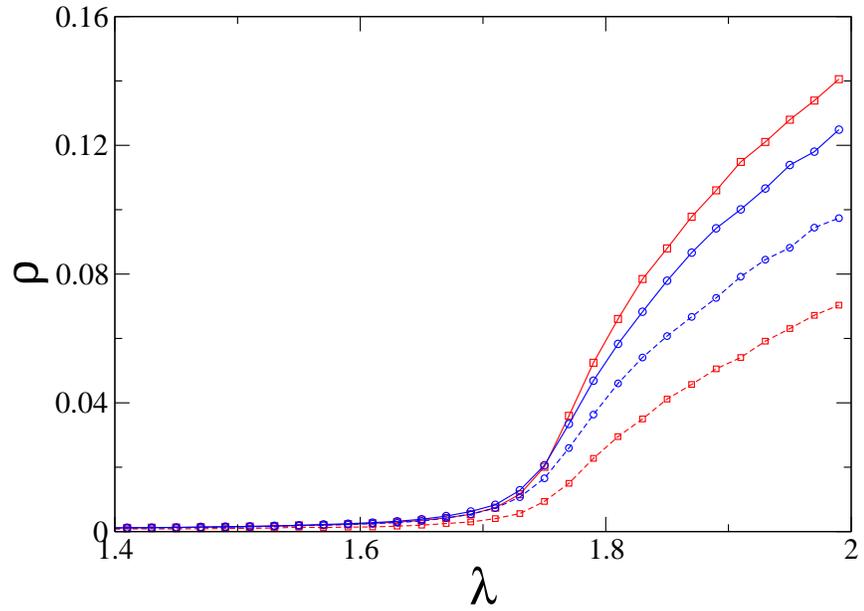}
\caption{\footnotesize{(Color online) Square lattice: quasistationary population densities $\rho_A$ (solid lines) and $\rho_B$ (dashed lines)
versus reproduction rate $\lambda$, for $\alpha=0.1$ (red) and $\alpha=0.5$ (blue). Linear system size: $L=80$.
}}
\label{cpfs2drho}
\end{figure}

We performed extensive Monte Carlo simulations of the CPFS on square lattices
using quasi-stationary (QS) simulations \cite{qssim1,qssim2}, restricting the dynamics to the subspace containing at least one individual of each species.
Our results show that the QS density of the slower species $B$ is always lower than that of the faster species $A$ (see Fig.~\ref{cpfs2drho}).

\begin{figure}[!hbt]
\includegraphics[clip,angle=0,width=.9\hsize]{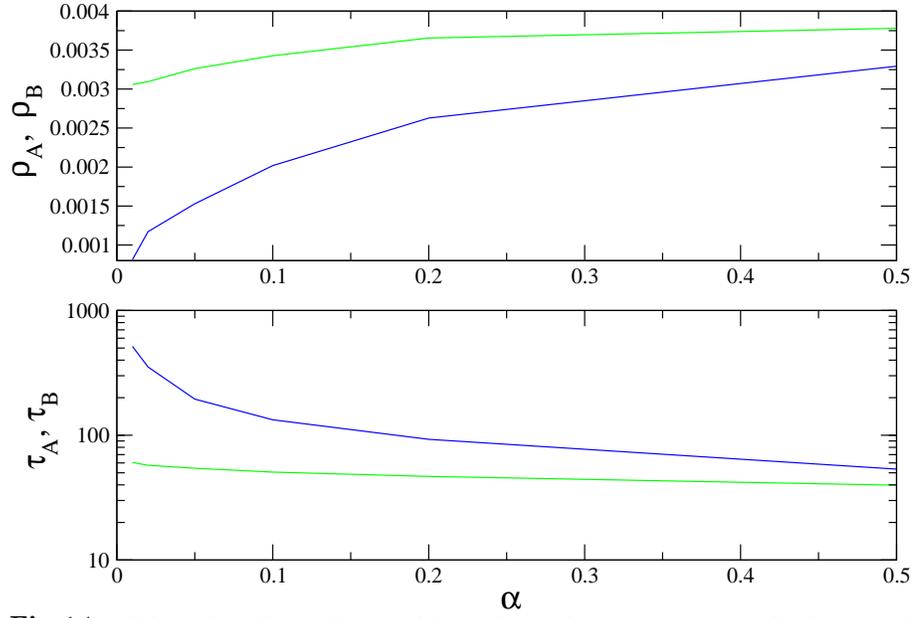}
\caption{\footnotesize{(Color online) Square lattice: QS population densities (upper panel) of species $A$ and $B$ and lifetime of the QS state (lower panel) for $\lambda=1.65$ and linear system size $L=80$.
}}
\label{cpfs2dalpha}
\end{figure}

\begin{figure}[!hbt]
\includegraphics[clip,angle=0,width=1.\hsize]{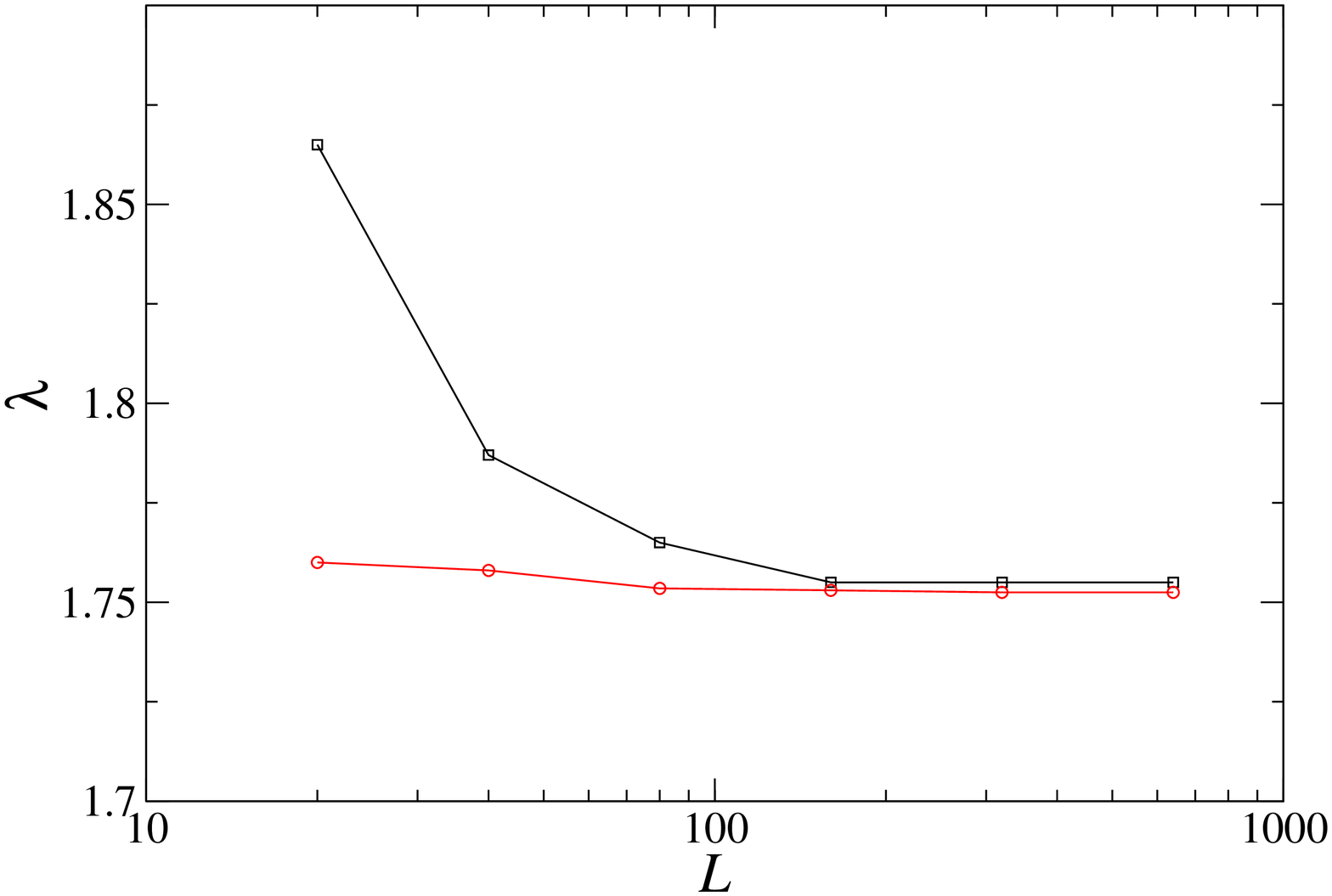}
\caption{\footnotesize{  (Color online) Boundaries of the SS regime in the $\lambda$ - $L$ plane for $\alpha=0.1$.  The lower boundary corresponds to
the pseudocritical point, $\lambda_c(L)$.}}
\label{cpfs2dss}
\end{figure}

The behavior of the QS densities and lifetimes is qualitatively similar to that on the complete graph.
We again find a range of values of $\alpha$ for which the slower (and less numerous) species is more likely to survive than the faster (and more populous) one (Fig.~\ref{cpfs2dalpha}).
This result is similar to the SS phenomenon \cite{ss}. In our model, however, SS only occurs for small system sizes. Figure~\ref{cpfs2dss} shows that the interval of $\lambda$ values over which SS is observed
decreases steadily with lattice size $L$.  The lower boundary in Fig.~\ref{cpfs2dss} which marks the active-aborning phase transition defines the pseudocritical points $\lambda_c(L)$, which converge to the critical value $\lambda_c$ when $L\to\infty$.

\subsection{Stochastic environment}

The effects of environmental or temporal disorder in systems exhibiting absorbing phase transitions have attracted interest recently \cite{jensen96,kamenev,munoz2011,neto,oliv-fiore16a,oliv-fiore16b,neto2}. In such
cases, the control parameter varies stochastically,
resulting in  temporarily active (ordered)
and absorbing (disordered) phases,
whose effects are more relevant at the emergence of the phase transition.
Since competitive dynamics often occurs
in stochastic environments \cite{kamenev, envnoise, borile2}, in this section we investigate the impact
of environmental noise on the CPFS on a complete graph. We assume that the
basic reproduction rate fluctuates according to
\begin{equation}
\lambda(t)=\lambda_0+X(t),
\end{equation}
where $\lambda_0$ is constant and $X(t)$ is the environmental noise, with $\langle X(t) \rangle = 0$.
In practice, we introduce the temporal disorder in the following way: at
each  time interval $t_i\le t\le t_i+\Delta t$ in the simulation,
$\lambda(t)$ assumes a new value, extracted from a uniform
distribution with mean $\lambda_0$ and width $\sigma$. More specifically,
$X(t)$ is evaluated using the formula $X(t)=(2\xi-1)\sigma$,
where $\xi$ is a random number drawn from a uniform probability distribution
 $\in [0,1]$  updated at fixed time intervals, $\Delta t$.

\begin{figure}[!hbt]
\includegraphics[clip,angle=0,width=1.0\hsize]{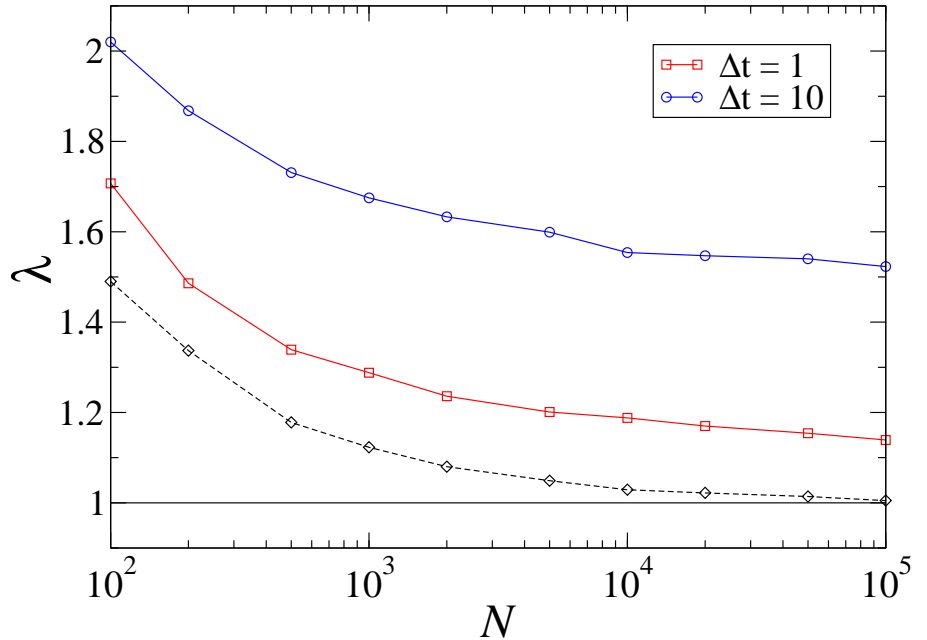}
\caption{\footnotesize{(Color online) CPFS on complete graph
with stochastic environment (QS simulations). Pseudocritical point $\lambda_c(N)$  (solid curve) and limit of the SS regime $\lambda^*(N)$ (circles) as function of system size $N$, for $\alpha=0.1$ and $\sigma=0.8$. Dashed lines: limit of the SS regime for the constant environment case.
}}
\label{cpfstemp}
\end{figure}

In Fig.~\ref{cpfstemp} we show the results of simulations with temporal disorder for noise intensity $\sigma=0.8$.
At $\lambda=\lambda_c=1.0$, both species go extinct, as expected. In the supercritical phase, the slower species is favored for
region $\lambda_c<\lambda<\lambda^*$, i.e., it has a higher survival probability than the faster one, despite being less populous. For $\lambda>\lambda*$ both species are equally fit, and have the same probability of survival.
Fig.~\ref{cpfstemp} shows that the
advantage of the slower species increases when $\Delta t$ (proportional to the noise autocorrelation time) is greater.
This is because increasing $\Delta t$,
periods with unfavorable values of $\lambda$ become longer, and the faster species is more
vulnerable to extinction during such periods.  We therefore conclude that the slower species again enjoys an advantage under temporal disorder.
In this case the advantage persists to much larger system sizes than in the absence of disorder (see Fig. 15 for comparison with the constant-environment case).

\section{Conclusion}

In summary, in this work we have investigated competitive advantage in the context of quasi-neutral systems, using
deterministic and stochastic mean-field descriptions, exact quasi-stationary distributions on a complete graph and
small rings, and simulations on a two-dimensional lattice.
We employ a two-species contact process (CP) in which the species differ only in the rates of their biological clocks, so that
their reproduction/death ratios are the same.  The CP features inter- and intraspecific competition for space.
In the macroscopic (MF) description, the slower species enjoys an advantage
in the form of a larger stationary population density for reproduction rates $\lambda$ close to (but greater than) the critical
value, and large initial population densities.  This persists under periodic (sinusoidal) variation of the reproduction rate, and
can lead to ``survival of the scarcer" (SS) in the presence of additive noise. In finite stochastic systems, with spatial
structure (lattices) or without it, (i.e., the complete graph), the slower species has a smaller population density than the
faster one, but can have a longer mean lifetime.
As a general tendency, the slow species is at an advantage in situations in which the overall population is decreasing, and vice-versa.
Since ability to weather the former, unfavorable, situation may be critical to survival, a slower species may have a long-term advantage over a
similar, faster one.
In finite stochastic systems this SS phenomenon is most prominent in small
systems near the critical point.  The advantage of the slower species vanishes for large reproduction rates and/or system sizes.
Our results suggest, however, that this advantage persists to large systems if the reproduction rate varies randomly, modeling
a stochastic environment.

Thus our results provide new insights into how stochasticity and competition interact to determine
survival in finite systems. We find that a species with a slower biological clock can be at an advantage
if resources are limited, and/or
if the system is subject to environmental variation. In the latter case (temporal disorder)
the advantage is more robust, persisting for large system sizes.  This advantage is also observed in the spatial model. In most cases, the faster species is
subject to a larger amount of demographic noise, and is more vulnerable to extinction. As a consequence, we observe that starting from total densities greater than the quasi stationary values,
the slow species usually lose less population and eventually dominate in the long term.
The importance of
incorporating demographic stochasticity into basic models of population genetics has been highlighted recently \cite{parsons,lin2012,parsons2,genetics,kogan,nelson,tarnita}. It was shown that if two phenotypes have equal deterministic fitness, but one is subject to a larger amount of demographic noise than the other, then the effect of this noise can induce a selective drift in favor of the phenotype experiencing less noise. In the same way, it is easier to invade a noisy population than a stable one. Therefore, a study of {\em  spatial} quasineutral models would be interesting in this context.
It could also help to understand the persistence of some organisms in antimicrobial therapy\cite{antimic1,antimic2}. Here, variation tends to accumulate in different individuals of evolutionary populations in situations where new variants are neutral or quasi-neutral; if the population size is not so large, genetic drift can be a stronger force than selection \cite{genetics,ohta1,ohta2}.

Finally, we note that, beyond ecology \cite{ecology}, competition for resources is of vital importance in epidemiology \cite{epidemics} and social sciences \cite{social}.
Possible extensions of the present work include the study of spatial inhomogeneities in the environment \cite{borile}. Such quenched disorder in quasi-neutral models could induce the appearance of Griffiths phases \cite{adr-dic,dcp1,dcp2} and refuges that could enhance the competitive edge of the faster species \cite{borile2}. A search for a general framework to study the ``slower is faster" effect \cite{helbing} (observed in pedestrian dynamics, vehicle traffic, logistics and social dynamics) is also an interesting research direction.

\nolinenumbers

\end{document}